\documentclass{article}
\usepackage{float}
\usepackage{braket}
\usepackage{multirow}
\usepackage{longtable}
\usepackage{subfigure}
\usepackage{authblk}
\usepackage{latexsym}
\usepackage{amssymb}
\usepackage{amsmath}
\usepackage{array}
\usepackage{graphicx}
\DeclareGraphicsExtensions{.pdf,.jpeg,.png}
\usepackage{epstopdf}
\usepackage{mathtools}
\usepackage{amsthm}
\usepackage{indentfirst}
\usepackage[misc]{ifsym}

\usepackage[numbers,sort&compress]{natbib}

\title{The masking condition for quantum state in two-dimensional Hilbert space}
\author[1]{Mei-Yi Wang}
\author[1]{Su-Juan Zhang \thanks{Corresponding author:zhangsj@stdu.edu.cn}}
\author[1]{Chen-Ming Bai}
\author[1]{Lu Liu}
\affil{Department of Mathematics and Physics, Shijiazhuang Tiedao University, Shijiazhuang 050043, China}

\date{}
\begin{document}

\maketitle

\begin{abstract}
This paper focuses on quantum information masking for quantum state in two-dimensional Hilbert space. We present a system of equations as the condition of quantum information masking. It is shown that quantum information contained in a single qubit state can be masked, if and only if the coefficients of quantum state satisfy the given system of equations. By observing the characteristics of non-orthogonal maskable quantum states, we obtain a related conclusion, namely, if two non-orthogonal two-qubit quantum states can mask a single qubit state, they have the same number of terms and the same basis. Finally, for maskable orthogonal quantum states, we analyze two special examples and give their images for an intuitive description.
\end{abstract}

\section{Introduction}

Quantum mechanics is one of the most important inventions in the field of physics in the 20th century \cite{ref1,ref2}. As a crucial research object of quantum mechanics, quantum information has very great significance. In quantum information theory, the evolution in a closed quantum system is unitary and linear \cite{ref3}. Therefore, unlike the classical world, there are some no-go theorems in the quantum world, such as the no-deleting theorem \cite{ref4,ref5,ref6}, the no-cloning theorem \cite{ref7,ref8,ref9}, the no-broadcasting theorem \cite{ref10,ref11,ref12} and so on. As one of no-go theorems, the no-masking theorem \cite{ref13} plays a vital role in the development of quantum information. Quantum information masking requires that the information in subsystems can be transferred into their composite quantum system by unitary operations, such that the final reduced states of any subsystems are identical \cite{ref14}. In other words, the initial information in the subsystems is hidden \cite{ref15}. This principle greatly improves the security of quantum communication \cite{ref16,ref17}.  As a consequence, the masking of quantum information can be used for quantum secret sharing \cite{ref18,ref19,ref20}. Besides, it also plays a major role in quantum teleportation \cite{ref21,ref22} and key distribution \cite{ref23}. Therefore, it is necessary to study the quantum states whether can be masked. At present, the research method of this problem is mainly based on the definition of quantum information masking.

In recent works, many scholars have come up with a lot of new discoveries about quantum masking. Modi et al. \cite{ref13} first proposed the no-masking theorem, namely, there is no unitary operator that can mask all pure states. Based on this theorem, Ghosh et al. \cite{ref24} considered two special kinds of quantum states with orthogonal and non-orthogonal basis states to obtain the masking conditions. Li and Wang \cite{ref25} presented some schemes different from error correction codes and showed that quantum states can be masked when more participants are allowed in the masking process. Then Han et al. \cite{ref26} further explored the relationship between quantum multipartite maskers (QMMs) and quantum error-correcting codes (QECCs). In addition, Ding et al. \cite{ref27} explicitly studied the structure of the set of maskable states and its relation to hyperdisks. Li and Modi \cite{ref28} studied probabilistic and approximate quantum information masking and generalised the no-masking theorem to allow for failure of the protocol. Lie et al. \cite{ref29} proposed the correlation between entropy and quantum information masking. They demonstrated that when a quantum state is a randomness source for a quantum masking process, the min-entropy of this state is the maximum number of qubits that can be masked. Liu et al. \cite{ref30} devised a photonic quantum information masking machine using time-correlated photons and investigated the properties of qubit masking. These above papers are important for the development of quantum information masking.

Here, we mainly study the masking of quantum state in two-dimensional Hilbert space. We express the masking conditions by the coefficients of quantum states. As a result, we obtain a system of equations as the masking conditions. Afterward, by observing some concrete maskable non-orthogonal quantum states, we find a common property of these states, i.e., they have the same number of terms and the same basis. Eventually, we analyze two special kinds of quantum states with orthogonal basis and draw some images for an intuitive description of these examples.

The organization of the paper is as follows. In Section \ref{section1}, we review the definition of quantum information masking and give the masking conditions of quantum state in two-dimensional Hilbert space. In Section \ref{section2}, firstly, we consider the masking of non-orthogonal quantum states and obtain a conclusion about their common property. Then, we analyse two examples of orthogonal quantum states and show their corresponding figures. Finally, we summarize the paper in Section \ref{section3}.

\section{No-masking theorem conditions}\label{section1}

In this section, we review the definition of masking \cite{ref13} and discuss the general masking condition of quantum state in $\mathbb{C}^{2}$.

Let $H_{X}$ be a $n$-dimensional Hilbert space associated with the system $X$. Suppose the states $|a_{k}\rangle_{A}$ in $H_{A}$ contain the quantum information. We say that the quantum information contained in states $|a_{k}\rangle_{A}$ can be masked, if and only if there is an operator $F$ that maps the states to $|\Psi_{k}\rangle_{AB}$ which belong to $H_{A}\otimes H_{B}$, such that the marginal states of $|\Psi_{k}\rangle_{AB}$ satisfy the following two conditions
\begin{align*}
\rho_{A}={\rm Tr}_{B}(|\Psi_{k}\rangle_{AB}\langle\Psi_{k}|),
\end{align*}
\begin{align*}
\rho_{B}={\rm Tr}_{A}(|\Psi_{k}\rangle_{AB}\langle\Psi_{k}|).
\end{align*}

That is to say we can't tell what the value of $k$ is or what information it carries by looking at the states. Since this is a physical process, we can also write it as $F:|a_{k}\rangle_{A}\otimes|b_{k}\rangle_{B}\rightarrow|\Psi_{k}\rangle_{AB}$, where the operator $F$ is called the masker and it is unitary, $|b_{k}\rangle_{B}\in H_{B}$. Moreover, in order to complete the difference between the dimensionality of two systems, there is an unitary operator $U$ acting on the system $A$ and $B$.

Assume that the quantum information in the state $|b\rangle=\alpha_{0}|0\rangle+\alpha_{1}|1\rangle$ can be masked. Then there exists two quantum states $|\Psi_{0}\rangle$ and $|\Psi_{1}\rangle$ which belong to $H_{A}\otimes H_{B}$, such that
\begin{align*}
|b\rangle=\alpha_{0}|0\rangle+\alpha_{1}|1\rangle\rightarrow|\Psi\rangle=\alpha_{0}|\Psi_{0}\rangle+\alpha_{1}|\Psi_{1}\rangle,
\end{align*}
where $|\alpha_{0}|^{2}+|\alpha_{1}|^{2}=1$.

For the subsystem $A$ and $B$, we take the partial trace respectively, therefore, we get
\begin{align}
\rho_{B}={\rm Tr}_{A}(|\Psi\rangle\langle\Psi|)=|&\alpha_{0}|^{2}{\rm Tr}_{A}(|\Psi_{0}\rangle\langle\Psi_{0}|)+|\alpha_{1}|^{2}{\rm Tr}_{A}(|\Psi_{1}\rangle\langle\Psi_{1}|){}\nonumber\\
{}+&\alpha_{0}\alpha_{1}^{\ast}{\rm Tr}_{A}(|\Psi_{0}\rangle\langle\Psi_{1}|)+\alpha_{0}^{\ast}\alpha_{1}{\rm Tr}_{A}(|\Psi_{1}\rangle\langle\Psi_{0}|),\label{1}\\
\nonumber\\
\rho_{A}={\rm Tr}_{B}(|\Psi\rangle\langle\Psi|)=|&\alpha_{0}|^{2}{\rm Tr}_{B}(|\Psi_{0}\rangle\langle\Psi_{0}|)+|\alpha_{1}|^{2}{\rm Tr}_{B}(|\Psi_{1}\rangle\langle\Psi_{1}|){}\nonumber\\
{}+&\alpha_{0}\alpha_{1}^{\ast}{\rm Tr}_{B}(|\Psi_{0}\rangle\langle\Psi_{1}|)+\alpha_{0}^{\ast}\alpha_{1}{\rm Tr}_{B}(|\Psi_{1}\rangle\langle\Psi_{0}|).\label{2}
\end{align}

By the definition of masking, the masking conditions are
\begin{align*}
\rho_{A}={\rm Tr}_{B}(|\Psi_{0}\rangle\langle\Psi_{0}|)={\rm Tr}_{B}(|\Psi_{1}\rangle\langle\Psi_{1}|)={\rm Tr}_{B}(|\Psi\rangle\langle\Psi|),\\
\\
\rho_{B}={\rm Tr}_{A}(|\Psi_{0}\rangle\langle\Psi_{0}|)={\rm Tr}_{A}(|\Psi_{1}\rangle\langle\Psi_{1}|)={\rm Tr}_{A}(|\Psi\rangle\langle\Psi|).
\end{align*}
In order to fulfill the above masking conditions, the cross-terms in Eq.(\ref{1}) and Eq.(\ref{2}) must be vanished, i.e.,
\begin{align}
\alpha_{0}\alpha_{1}^{\ast}{\rm Tr}_{A}(|\Psi_{0}\rangle\langle\Psi_{1}|)+\alpha_{0}^{\ast}\alpha_{1}{\rm Tr}_{A}(|\Psi_{1}\rangle\langle\Psi_{0}|)=0,\label{3}\\
\nonumber\\
\alpha_{0}\alpha_{1}^{\ast}{\rm Tr}_{B}(|\Psi_{0}\rangle\langle\Psi_{1}|)+\alpha_{0}^{\ast}\alpha_{1}{\rm Tr}_{B}(|\Psi_{1}\rangle\langle\Psi_{0}|)=0.
\end{align}

In this paper, we mainly discuss the masking of quantum state in two-dimensional Hilbert space. Without loss of generality, suppose that
\begin{align*}
|\Psi_{0}\rangle&=a_{0}|00\rangle+a_{1}|01\rangle+a_{2}|10\rangle+a_{3}|11\rangle,\\\\
|\Psi_{1}\rangle&=b_{0}|00\rangle+b_{1}|01\rangle+b_{2}|10\rangle+b_{3}|11\rangle,
\end{align*}
where $\sum_{i=0}^{3}|a_{i}|^{2}=\sum_{i=0}^{3}|b_{i}|^{2}=1$.

When ${\rm Tr}_{x}(|\Psi_{0}\rangle\langle\Psi_{0}|)={\rm Tr}_{x}(|\Psi_{1}\rangle\langle\Psi_{1}|)$, where $x\in\{A,B\}$, the coefficients of $|\Psi_{0}\rangle$ and $|\Psi_{1}\rangle$ should meet the following system of equations
\begin{align}
\begin{split}
\left \{
\begin{array}{llllll}
|a_{0}|^{2}+|a_{1}|^{2}-|b_{0}|^{2}-|b_{1}|^{2}=0\\\\
|a_{0}|^{2}+|a_{2}|^{2}-|b_{0}|^{2}-|b_{2}|^{2}=0\\\\
|a_{1}|^{2}+|a_{3}|^{2}-|b_{1}|^{2}-|b_{3}|^{2}=0\\\\
|a_{2}|^{2}+|a_{3}|^{2}-|b_{2}|^{2}-|b_{3}|^{2}=0\\\\
a_{0}a_{1}^{*}+a_{2}a_{3}^{*}-b_{0}b_{1}^{*}-b_{2}b_{3}^{*}=0\\\\
a_{0}a_{2}^{*}+a_{1}a_{3}^{*}-b_{0}b_{2}^{*}-b_{1}b_{3}^{*}=0.
\end{array}
\right.
\end{split}
\label{4}
\end{align}

In addition, we take the partial trace respect to $A$ for $|\Psi_{0}\rangle\langle\Psi_{1}|$ and $|\Psi_{1}\rangle\langle\Psi_{0}|$, then we gain
\begin{align}
{\rm Tr}_{A}(|\Psi_{0}\rangle\langle\Psi_{1}|)=(&a_{0}b_{0}^{\ast}+a_{2}b_{2}^{\ast})|0\rangle\langle0|+(a_{0}b_{1}^{\ast}+a_{2}b_{3}^{\ast})|0\rangle\langle1|{}\nonumber\\
{}+(&a_{1}b_{0}^{\ast}+a_{3}b_{2}^{\ast})|1\rangle\langle0|+(a_{1}b_{1}^{\ast}+a_{3}b_{3}^{\ast})|1\rangle\langle1|,\label{5}\\
\nonumber\\
{\rm Tr}_{A}(|\Psi_{1}\rangle\langle\Psi_{0}|)=(&a_{0}^{\ast}b_{0}+a_{2}^{\ast}b_{2})|0\rangle\langle0|+(a_{1}^{\ast}b_{0}+a_{3}^{\ast}b_{2})|0\rangle\langle1|{}\nonumber\\
{}+(&a_{0}^{\ast}b_{1}+a_{2}^{\ast}b_{3})|1\rangle\langle0|+(a_{1}^{\ast}b_{1}+a_{3}^{\ast}b_{3})|1\rangle\langle1|.\label{6}
\end{align}

For the convenience of discussion, we denote
\begin{align*}
A=a_{0}b_{0}^{\ast}+a_{2}b_{2}^{\ast},\qquad B=a_{0}b_{1}^{\ast}+a_{2}b_{3}^{\ast},\\
C=a_{1}b_{0}^{\ast}+a_{3}b_{2}^{\ast},\qquad D=a_{1}^{\ast}b_{1}+a_{3}^{\ast}b_{3}.
\end{align*}

Substituting Eq.(\ref{5}) and Eq.(\ref{6}) into Eq.(\ref{3}), we obtain
\begin{align}
&(\alpha_{1}\alpha_{2}^{*}A+\alpha_{1}^{*}\alpha_{2}A^{*})|0\rangle\langle0|+(\alpha_{1}\alpha_{2}^{*}B+\alpha_{1}^{*}\alpha_{2}C^{*})|0\rangle\langle1|{}\nonumber\\
{}+&(\alpha_{1}\alpha_{2}^{*}C+\alpha_{1}^{*}\alpha_{2}B^{*})|1\rangle\langle0|+(\alpha_{1}\alpha_{2}^{*}D+\alpha_{1}^{*}\alpha_{2}D^{*})|1\rangle\langle1|=0.\nonumber
\end{align}

As a result, the conditions for the establishment of the above formula can be expressed as
\begin{align}
\begin{split}
\left \{
\begin{array}{llll}
\alpha_{1}\alpha_{2}^{*}A+\alpha_{1}^{*}\alpha_{2}A^{*}=0\nonumber\\\\
\alpha_{1}\alpha_{2}^{*}B+\alpha_{1}^{*}\alpha_{2}C^{*}=0\nonumber\\\\
\alpha_{1}\alpha_{2}^{*}C+\alpha_{1}^{*}\alpha_{2}B^{*}=0\nonumber\\\\
\alpha_{1}\alpha_{2}^{*}D+\alpha_{1}^{*}\alpha_{2}D^{*}=0.\nonumber
\end{array}
\right.
\end{split}
\end{align}

This set of conditions can be farther simplified as
\begin{align}
\begin{split}
\left \{
\begin{array}{lll}
{\rm Re}(\alpha_{1}\alpha_{2}^{*}A)=0\\\\
{\rm Re}(\alpha_{1}\alpha_{2}^{*}D)=0\\\\
\alpha_{1}\alpha_{2}^{*}B+\alpha_{1}^{*}\alpha_{2}C^{*}=0.
\end{array}
\right.
\end{split}
\label{7}
\end{align}

Similar to the above method, for the subsystem $B$, we acquire that
\begin{align}
\begin{split}
\left \{
\begin{array}{lll}
{\rm Re}(\alpha_{1}\alpha_{2}^{*}A^{'})=0\\\\
{\rm Re}(\alpha_{1}\alpha_{2}^{*}D^{'})=0\\\\
\alpha_{1}\alpha_{2}^{*}B^{'}+\alpha_{1}^{*}\alpha_{2}C^{'*}=0,
\end{array}
\right.
\end{split}
\label{8}
\end{align}
where
\begin{align*}
A^{'}=a_{0}b_{0}^{\ast}+a_{1}b_{1}^{\ast},\qquad B^{'}=a_{0}b_{2}^{\ast}+a_{1}b_{3}^{\ast},\\
C^{'}=a_{2}b_{0}^{\ast}+a_{3}b_{1}^{\ast},\qquad D^{'}=a_{2}b_{2}^{\ast}+a_{3}b_{3}^{\ast}.
\end{align*}

In summary, the masking conditions can be represented by the coefficients of $|\Psi\rangle$, $|\Psi_{0}\rangle$ and $|\Psi_{1}\rangle$, i.e., when their coefficients meet the requirements of Eq.(\ref{4}), Eq.(\ref{7}) and Eq.(\ref{8}) at the same time, a single qubit state $|b\rangle=\alpha_{0}|0\rangle+\alpha_{1}|1\rangle$ can be masked.

\section{Masking of two-qubit quantum states}\label{section2}

Here, we consider some non-orthogonal quantum states and orthogonal quantum states respectively, then we get some corresponding conclusions as follows.

\subsection{Masking of non-orthogonal quantum states}

By observing some maskable non-orthogonal quantum states in $\mathbb{C}^{2}$, we obtain the result that if two non-orthogonal quantum states can mask quantum information contained in a single qubit state, they have the same number of terms and the same basis. Below, we present the analysis process of this result.

Assume that $|\Psi_{0}\rangle$ and $|\Psi_{1}\rangle$ are two non-orthogonal quantum states in $\mathbb{C}^{2}\otimes\mathbb{C}^{2}$, and $|\Psi\rangle=\alpha_{0}|\Psi_{0}\rangle+\alpha_{1}|\Psi_{1}\rangle$ can mask the quantum information in the state $|b\rangle=\alpha_{0}|0\rangle+\alpha_{1}|1\rangle$, where $\sum_{i=0}^{1}|\alpha_{i}|^{2}=1$. Hence, we have
\begin{align*}
{\rm Tr}_{A}(|\Psi\rangle\langle\Psi|)={\rm Tr}_{A}(|\Psi_{0}\rangle\langle\Psi_{0}|)={\rm Tr}_{A}(|\Psi_{1}\rangle\langle\Psi_{1}|),\\\\
{\rm Tr}_{B}(|\Psi\rangle\langle\Psi|)={\rm Tr}_{B}(|\Psi_{0}\rangle\langle\Psi_{0}|)={\rm Tr}_{B}(|\Psi_{1}\rangle\langle\Psi_{1}|).
\end{align*}

We discuss four possible cases of $|\Psi_{0}\rangle$ respectively to make an analysis as follows.

\textbf{Case 1.} When $|\Psi_{0}\rangle$ consists of one term, i.e., $|\Psi_{0}\rangle=|i_{1}i_{2}\rangle$, where $i_{1}$, $i_{2}\in\{0, 1\}$. Assume that $|\Psi_{1}\rangle=b|i_{1}i_{2}\rangle+c|j_{1}j_{2}\rangle$, where $b\neq0$, $c\neq0$, $|b|^{2}+|c|^{2}=1$ and $i_{1}$, $i_{2}$, $j_{1}$, $j_{2}\in\{0,1\}$. The specific classification of $|\Psi_{0}\rangle$, $|\Psi_{1}\rangle$ and the corresponding masking situation are shown in Table \ref{table:1}.
\begin{table}[htbp]
  \centering
  \caption{$|\Psi_{0}\rangle$ consists of one term(The duplicate parts have been deleted)}
  \label{table:1}
  \begin{tabular}{|c|c|c|c|}
  \hline
  $|\Psi_{0}\rangle=|i_{1}i_{2}\rangle$ & \multicolumn{2}{c|}{$|\Psi_{1}\rangle=b|i_{1}i_{2}\rangle+c|j_{1}j_{2}\rangle$} & \multirow{2}{*}{Masking or not}\\
  \cline{1-3}
  $|i_{1}i_{2}\rangle$ & $|i_{1}i_{2}\rangle$ & $|j_{1}j_{2}\rangle$ & \\
  \hline
  \multirow{4}{*}{$|00\rangle$} & $|00\rangle$ & $|00\rangle$ & $\surd$\\
  \cline{2-4}
  & $|00\rangle$ & $|01\rangle$ & $\times$\\
  \cline{2-4}
  & $|00\rangle$ & $|10\rangle$ & $\times$\\
  \cline{2-4}
  & $|00\rangle$ & $|11\rangle$ & $\times$\\
  \hline
  \multirow{4}{*}{$|01\rangle$} & $|01\rangle$ & $|00\rangle$ & $\times$\\
  \cline{2-4}
  & $|01\rangle$ & $|01\rangle$ & $\surd$\\
  \cline{2-4}
  & $|01\rangle$ & $|10\rangle$ & $\times$\\
  \cline{2-4}
  & $|01\rangle$ & $|11\rangle$ & $\times$\\
  \hline
  \multirow{4}{*}{$|10\rangle$} & $|10\rangle$ & $|00\rangle$ & $\times$\\
  \cline{2-4}
  & $|10\rangle$ & $|01\rangle$ & $\times$\\
  \cline{2-4}
  & $|10\rangle$ & $|10\rangle$ & $\surd$\\
  \cline{2-4}
  & $|10\rangle$ & $|11\rangle$ & $\times$\\
  \hline
  \multirow{4}{*}{$|11\rangle$} & $|11\rangle$ & $|00\rangle$ & $\times$\\
  \cline{2-4}
  & $|11\rangle$ & $|01\rangle$ & $\times$\\
  \cline{2-4}
  & $|11\rangle$ & $|10\rangle$ & $\times$\\
  \cline{2-4}
  & $|11\rangle$ & $|11\rangle$ & $\surd$\\
  \hline
  \end{tabular}
\end{table}

According to Table \ref{table:1}, we take the case of $i_{1}=j_{1}$ and $i_{2}\neq j_{2}$ as an example. It can be seen that
\begin{align*}
{\rm Tr}_{A}(|\Psi_{0}\rangle\langle\Psi_{0}|)&=|i_{2}\rangle\langle i_{2}|,\\\\
{\rm Tr}_{A}(|\Psi_{1}\rangle\langle\Psi_{1}|)&=|b|^{2}|i_{2}\rangle\langle i_{2}|+bc^{*}|i_{2}\rangle\langle j_{2}|+b^{*}c|j_{2}\rangle\langle i_{2}|+|c|^{2}|j_{2}\rangle\langle j_{2}|.
\end{align*}
Since $b$, $c\neq0$, then ${\rm Tr}_{A}(|\Psi_{0}\rangle\langle\Psi_{0}|)\neq {\rm Tr}_{A}(|\Psi_{1}\rangle\langle\Psi_{1}|)$, which contradicts the masking condition.

\textbf{Case 2.} Suppose $|\Psi_{0}\rangle=a_{0}|i_{1}i_{2}\rangle+a_{1}|j_{1}j_{2}\rangle$, where $a_{0}$, $a_{1}\neq0$, $|a_{0}|^{2}+|a_{1}|^{2}=1$ and $i_{1}$, $i_{2}$, $j_{1}$, $j_{2}\in\{0, 1\}$.

Assume $|\Psi_{1}\rangle=b_{0}|i_{1}i_{2}\rangle+b_{1}|k_{1}k_{2}\rangle+b_{2}|l_{1}l_{2}\rangle$, where $\sum_{i=0}^{2}|b_{i}|^{2}=1$, $b_{0}$, $b_{1}$, $b_{2}\neq0$ and $i_{1}$, $i_{2}$, $k_{1}$, $k_{2}$, $l_{1}$, $l_{2}\in\{0, 1\}$. By classifying and discussing the relationship among $i_{1}$, $i_{2}$, $j_{1}$, $j_{2}$, $k_{1}$, $k_{2}$, $l_{1}$, $l_{2}$, we take the partial trace of $|\Psi_{0}\rangle$ and $|\Psi_{1}\rangle$ respect to both system $A$ and system $B$, through which we can know whether $|\Psi_{0}\rangle$ and $|\Psi_{1}\rangle$ can mask the single qubit state. The corresponding situation is shown in Table \ref{table:2}.

\begin{table}[htbp]
  \centering
  \caption{$|\Psi_{0}\rangle$ consists of two terms(The duplicate parts have been deleted)}
  \label{table:2}
    \resizebox{\textwidth}{!}{
  \begin{tabular}{|c|c|c|c|c|c|}
  \hline
  \multicolumn{2}{| c |}{$|\Psi_{0}\rangle=a_{0}|i_{1}i_{2}\rangle+a_{1}|j_{1}j_{2}\rangle$} & \multicolumn{3}{ c |}{$|\Psi_{1}\rangle=b_{0}|i_{1}i_{2}\rangle+b_{1}|k_{1}k_{2}\rangle+b_{2}|l_{1}l_{2}\rangle$} & \multirow{2}{*}{Masking or not}\\
  \cline{1-5}
  $|i_{1}i_{2}\rangle$ & $|j_{1}j_{2}\rangle$ & $|i_{1}i_{2}\rangle$ & $|k_{1}k_{2}\rangle$ & $|l_{1}l_{2}\rangle$ & \\
  \hline
  \multirow{7}{*}{$|00\rangle$} & \multirow{7}{*}{$|01\rangle$} & $|00\rangle$ & $|00\rangle$ & $|00\rangle$ & $\times$\\
  \cline{3-6}
  & & $|00\rangle$ & $|00\rangle$ & $|01\rangle$ & $\surd$\\
  \cline{3-6}
  & & $|00\rangle$ & $|00\rangle$ & $|10\rangle$ & $\times$\\
  \cline{3-6}
  & & $|00\rangle$ & $|00\rangle$ & $|11\rangle$ & $\times$\\
  \cline{3-6}
  & & $|00\rangle$ & $|01\rangle$ & $|10\rangle$ & $\times$\\
  \cline{3-6}
  & & $|00\rangle$ &$ |01\rangle$ & $|11\rangle$ & $\times$\\
  \cline{3-6}
  & & $|00\rangle$ & $|10\rangle$ & $|11\rangle$ & $\times$\\
  \hline
  \multirow{7}{*}{$|00\rangle$} & \multirow{7}{*}{$|10\rangle$} & $|00\rangle$ & $|00\rangle$ & $|00\rangle$ & $\times$\\
  \cline{3-6}
  & & $|00\rangle$ & $|00\rangle$ & $|01\rangle$ & $\times$\\
  \cline{3-6}
  & & $|00\rangle$ & $|00\rangle$ & $|10\rangle$ & $\surd$\\
  \cline{3-6}
  & & $|00\rangle$ & $|00\rangle$ & $|11\rangle$ & $\times$\\
  \cline{3-6}
  & & $|00\rangle$ & $|01\rangle$ & $|10\rangle$ & $\times$\\
  \cline{3-6}
  & & $|00\rangle$ & $|01\rangle$ & $|11\rangle$ & $\times$\\
  \cline{3-6}
  & & $|00\rangle$ & $|10\rangle$ & $|11\rangle$ & $\times$\\
  \hline
  \multirow{7}{*}{$|00\rangle$} & \multirow{7}{*}{$|11\rangle$} & $|00\rangle$ & $|00\rangle$ & $|00\rangle$ & $\times$\\
  \cline{3-6}
  & & $|00\rangle$ & $|00\rangle$ & $|01\rangle$ & $\times$\\
  \cline{3-6}
  & & $|00\rangle$ & $|00\rangle$ & $|10\rangle$ & $\times$\\
  \cline{3-6}
  & & $|00\rangle$ & $|00\rangle$ & $|11\rangle$ & $\surd$\\
  \cline{3-6}
  & & $|00\rangle$ & $|01\rangle$ & $|10\rangle$ & $\times$\\
  \cline{3-6}
  & & $|00\rangle$ & $|01\rangle$ & $|11\rangle$ & $\times$\\
  \cline{3-6}
  & & $|00\rangle$ & $|10\rangle$ & $|11\rangle$ & $\times$\\
  \hline
  \multirow{7}{*}{$|01\rangle$} & \multirow{7}{*}{$|10\rangle$} & $|01\rangle$ & $|00\rangle$ & $|00\rangle$ & $\times$\\
  \cline{3-6}
  & & $|01\rangle$ & $|00\rangle$ & $|10\rangle$ & $\times$\\
  \cline{3-6}
  & & $|01\rangle$ & $|00\rangle$ & $|11\rangle$ & $\times$\\
  \cline{3-6}
  & & $|01\rangle$ & $|01\rangle$ & $|01\rangle$ & $\times$\\
  \cline{3-6}
  & & $|01\rangle$ & $|01\rangle$ & $|10\rangle$ & $\surd$\\
  \cline{3-6}
  & & $|01\rangle$ & $|01\rangle$ & $|11\rangle$ & $\times$\\
  \cline{3-6}
  & & $|01\rangle$ & $|10\rangle$ & $|11\rangle$ & $\times$\\
  \hline
  \multirow{7}{*}{$|01\rangle$} & \multirow{7}{*}{$|11\rangle$} & $|01\rangle$ & $|00\rangle$ & $|00\rangle$ & $\times$\\
  \cline{3-6}
  & & $|01\rangle$ & $|00\rangle$& $|10\rangle$ & $\times$\\
  \cline{3-6}
  & & $|01\rangle$ & $|00\rangle$ & $|11\rangle$ & $\times$\\
  \cline{3-6}
  & & $|01\rangle$ & $|01\rangle$ & $|01\rangle$ & $\times$\\
  \cline{3-6}
  & & $|01\rangle$ & $|01\rangle$ & $|10\rangle$ & $\times$\\
  \cline{3-6}
  & & $|01\rangle$ & $|01\rangle$ & $|11\rangle$ & $\surd$\\
  \cline{3-6}
  & & $|01\rangle$ & $|10\rangle$ & $|11\rangle$ & $\times$\\
  \hline
  \multirow{7}{*}{$|10\rangle$} & \multirow{7}{*}{$|11\rangle$} & $|10\rangle$ & $|00\rangle$ & $|00\rangle$ & $\times$\\
  \cline{3-6}
  & & $|10\rangle$ & $|00\rangle$ & $|01\rangle$ & $\times$\\
  \cline{3-6}
  & & $|10\rangle$ & $|00\rangle$ & $|11\rangle$ & $\times$\\
  \cline{3-6}
  & & $|10\rangle$ & $|01\rangle$ & $|01\rangle$ & $\times$\\
  \cline{3-6}
  & & $|10\rangle$ & $|01\rangle$ & $|11\rangle$ & $\times$\\
  \cline{3-6}
  & & $|10\rangle$ & $|10\rangle$ & $|10\rangle$ & $\times$\\
  \cline{3-6}
  & & $|10\rangle$ & $|10\rangle$ & $|11\rangle$ & $\surd$\\
  \hline
  \end{tabular}}
\end{table}

Select the case of $|\Psi_{0}\rangle=a_{0}|00\rangle+a_{1}|01\rangle$ and $|\Psi_{1}\rangle=b_{0}|00\rangle+b_{1}|01\rangle+b_{2}|10\rangle$.

We obtain that
\begin{align*}
\qquad{\rm Tr}_{B}(|\Psi_{0}\rangle\langle\Psi_{0}|)&=|0\rangle\langle0|,\\\\
\qquad{\rm Tr}_{B}(|\Psi_{1}\rangle\langle\Psi_{1}|)&=(|b_{0}|^{2}+|b_{1}|^{2})|0\rangle\langle0|+b_{0}b_{2}^{*}|0\rangle\langle1|+b_{0}^{*}b_{2}|1
\rangle\langle0|+|b_{1}|^{2}|1\rangle\langle1|,
\end{align*}
where $b_{0}$, $b_{1}$, $b_{2}\neq0$. Then ${\rm Tr}_{B}(|\Psi_{0}\rangle\langle\Psi_{0}|)\neq {\rm Tr}_{B}(|\Psi_{1}\rangle\langle\Psi_{1}|)$, which contradicts the condition of masking.

\textbf{Case 3.} When $|\Psi_{0}\rangle$ is constitutive of three terms, namely, $|\Psi_{0}\rangle=a_{0}|i_{1}i_{2}\rangle+a_{1}|j_{1}j_{2}\rangle+a_{2}|k_{1}k_{2}\rangle$, where $a_{0}$, $a_{1}$, $a_{2}\neq0$, $\sum_{i=0}^{2}|a_{i}|^{2}=1$ and $i_{1}$, $i_{2}$, $j_{1}$, $j_{2}$, $k_{1}$, $k_{2}\in\{0, 1\}$.

Suppose that $|\Psi_{1}\rangle=b_{0}|i_{1}i_{2}\rangle+b_{1}|l_{1}l_{2}\rangle+b_{2}|m_{1}m_{2}\rangle+b_{3}|n_{1}n_{2}\rangle$, where $b_{0}$, $b_{1}$, $b_{2}$, $b_{3}\neq0$, $\sum_{i=0}^{3}|b_{i}|^{2}=1$ and $i_{1}$, $i_{2}$, $l_{1}$, $l_{2}$, $m_{1}$, $m_{2}$, $n_{1}$, $n_{2}\in\{0, 1\}$. We get the Table 3, similar to the above.
\begin{table}[htbp]
  \centering
  \caption{$|\Psi_{0}\rangle$ consists of three terms(The duplicate parts have been deleted)}
  \label{table:3}
   \resizebox{\textwidth}{!}{
  \begin{tabular}{|c|c|c|c|c|c|c|c|}
  \hline
  \multicolumn{3}{|c|}{$|\Psi_{0}\rangle=a_{0}|i_{1}i_{2}\rangle+a_{1}|j_{1}j_{2}\rangle+a_{2}|k_{1}k_{2}\rangle$} & \multicolumn{4}{c|}{$|\Psi_{1}\rangle=b_{0}|i_{1}i_{2}\rangle+b_{1}|l_{1}l_{2}\rangle+b_{2}|m_{1}m_{2}\rangle+b_{3}|n_{1}n_{2}\rangle$} & \multirow{2}{*}{Masking or not}\\
  \cline{1-7}
  $|i_{1}i_{2}\rangle$ & $|j_{1}j_{2}\rangle$ & $|k_{1}k_{2}\rangle$ & $|i_{1}i_{2}\rangle$ & $|l_{1}l_{2}\rangle$ & $|m_{1}m_{2}\rangle$ & $|n_{1}n_{2}\rangle$ & \\
  \hline
  \multirow{8}{*}{$|00\rangle$} & \multirow{8}{*}{$|01\rangle$} & \multirow{8}{*}{$|10\rangle$} & $|00\rangle$ & $|00\rangle$ & $|00\rangle$ & $|00\rangle$ & $\times$\\
  \cline{4-8}
  & & & $|00\rangle$ & $|00\rangle$ & $|00\rangle$ & $|01\rangle$ & $\times$\\
  \cline{4-8}
  & & & $|00\rangle$ & $|00\rangle$ & $|00\rangle$ & $|10\rangle$ & $\times$\\
  \cline{4-8}
  & & & $|00\rangle$ & $|00\rangle$ & $|00\rangle$ & $|11\rangle$ & $\times$\\
  \cline{4-8}
  & & & $|00\rangle$ & $|00\rangle$ & $|01\rangle$ & $|10\rangle$ & $\surd$\\
  \cline{4-8}
  & & & $|00\rangle$ & $|00\rangle$ & $|01\rangle$ & $|11\rangle$ & $\times$\\
  \cline{4-8}
  & & & $|00\rangle$ & $|00\rangle$ & $|10\rangle$ & $|11\rangle$ & $\times$\\
  \cline{4-8}
  & & & $|00\rangle$ & $|01\rangle$ & $|10\rangle$ & $|11\rangle$ & $\times$\\
  \hline
  \multirow{8}{*}{$|00\rangle$} & \multirow{8}{*}{$|01\rangle$} & \multirow{8}{*}{$|11\rangle$} & $|00\rangle$ & $|00\rangle$ & $|00\rangle$ & $|00\rangle$ & $\times$\\
  \cline{4-8}
  & & & $|00\rangle$ & $|00\rangle$ & $|00\rangle$ & $|01\rangle$ & $\times$\\
  \cline{4-8}
  & & & $|00\rangle$ & $|00\rangle$ & $|00\rangle$ & $|10\rangle$ & $\times$\\
  \cline{4-8}
  & & & $|00\rangle$ & $|00\rangle$ & $|00\rangle$ & $|11\rangle$ & $\times$\\
  \cline{4-8}
  & & & $|00\rangle$ & $|00\rangle$ & $|01\rangle$ & $|10\rangle$ & $\times$\\
  \cline{4-8}
  & & & $|00\rangle$ & $|00\rangle$ & $|01\rangle$ & $|11\rangle$ & $\surd$\\
  \cline{4-8}
  & & & $|00\rangle$ & $|00\rangle$ & $|10\rangle$ & $|11\rangle$ & $\times$\\
  \cline{4-8}
  & & & $|00\rangle$ & $|01\rangle$ & $|10\rangle$ & $|11\rangle$ & $\times$\\
  \hline
  \multirow{8}{*}{$|00\rangle$} & \multirow{8}{*}{$|10\rangle$} & \multirow{8}{*}{$|11\rangle$} & $|00\rangle$ & $|00\rangle$ & $|00\rangle$ & $|00\rangle$ & $\times$\\
  \cline{4-8}
  & & & $|00\rangle$ & $|00\rangle$ & $|00\rangle$ & $|01\rangle$ & $\times$\\
  \cline{4-8}
  & & & $|00\rangle$ & $|00\rangle$ & $|00\rangle$ & $|10\rangle$ & $\times$\\
  \cline{4-8}
  & & & $|00\rangle$ & $|00\rangle$ & $|00\rangle$ & $|11\rangle$ & $\times$\\
  \cline{4-8}
  & & & $|00\rangle$ & $|00\rangle$ & $|01\rangle$ & $|10\rangle$ & $\times$\\
  \cline{4-8}
  & & & $|00\rangle$ & $|00\rangle$ & $|01\rangle$ & $|11\rangle$ & $\times$\\
  \cline{4-8}
  & & & $|00\rangle$ & $|00\rangle$ & $|10\rangle$ & $|11\rangle$ & $\surd$\\
  \cline{4-8}
  & & & $|00\rangle$ & $|01\rangle$ & $|10\rangle$ & $|11\rangle$ & $\times$\\
  \hline
  \multirow{8}{*}{$|01\rangle$} & \multirow{8}{*}{$|10\rangle$} & \multirow{8}{*}{$|11\rangle$} & $|01\rangle$ & $|01\rangle$ & $|01\rangle$ & $|00\rangle$ & $\times$\\
  \cline{4-8}
  & & & $|01\rangle$ & $|01\rangle$ & $|01\rangle$ & $|01\rangle$ & $\times$\\
  \cline{4-8}
  & & & $|01\rangle$ & $|01\rangle$ & $|01\rangle$ & $|10\rangle$ & $\times$\\
  \cline{4-8}
  & & & $|01\rangle$ & $|01\rangle$ & $|01\rangle$ & $|11\rangle$ & $\times$\\
  \cline{4-8}
  & & & $|01\rangle$ & $|01\rangle$ & $|00\rangle$ & $|10\rangle$ & $\times$\\
  \cline{4-8}
  & & & $|01\rangle$ & $|01\rangle$ & $|00\rangle$ & $|11\rangle$ & $\times$\\
  \cline{4-8}
  & & & $|01\rangle$ & $|01\rangle$ & $|10\rangle$ & $|11\rangle$ & $\surd$\\
  \cline{4-8}
  & & & $|01\rangle$ & $|10\rangle$ & $|00\rangle$ & $|11\rangle$ & $\times$\\
  \hline
  \end{tabular}}

\end{table}

We choose the case of $|\Psi_{0}\rangle=a_{0}|00\rangle+a_{1}|01\rangle+a_{2}|10\rangle$ and $|\Psi_{1}\rangle=b_{0}|00\rangle+b_{1}|01\rangle+b_{2}|10\rangle+b_{3}|11\rangle$ from Table \ref{table:3} and get
\begin{align*}
\qquad\quad{\rm Tr}_{A}(|\Psi_{0}\rangle\langle\Psi_{0}|)=(|&a_{0}|^{2}\!+\!|a_{2}|^{2})|0\rangle\langle0|\!+\!|a_{1}|^{2}|1\rangle\langle1|\!+\!a_{0}a_{1}^{*}|0\rangle\langle1|\!+\!a_{0}^{*}a_{1}|1\rangle\langle0|,\nonumber\\\\
\qquad\quad{\rm Tr}_{A}(|\Psi_{1}\rangle\langle\Psi_{1}|)=(|&b_{0}|^{2}+|b_{2}|^{2})|0\rangle\langle0|+(|b_{1}|^{2}+|b_{3}|^{2})|1\rangle\langle1|\nonumber\\+(&b_{0}b_{1}^{*}+b_{2}b_{3}^{*})|0\rangle\langle1|+b_{0}^{*}b_{1}|1
\rangle\langle0|.\nonumber
\end{align*}

Since ${\rm Tr}_{A}(|\Psi_{0}\rangle\langle\Psi_{0}|)={\rm Tr}_{A}(|\Psi_{1}\rangle\langle\Psi_{1}|)$, it can be seen that
\begin{align}
\begin{split}
\left \{
\begin{array}{llll}
|a_{0}|^{2}+|a_{2}|^{2}-|b_{0}|^{2}-|b_{2}|^{2}=0\nonumber\\\\
|a_{1}|^{2}-|b_{1}|^{2}-|b_{3}|^{2}=0\nonumber\\\\
a_{0}a_{1}^{*}-b_{0}b_{1}^{*}-b_{2}b_{3}^{*}=0\nonumber\\\\
a_{0}^{*}a_{1}-b_{0}^{*}b_{1}=0.\nonumber
\end{array}
\right.
\end{split}
\end{align}
Then we get $b_{2}b_{3}^{*}=0$, i.e., $b_{2}=0$ or $b_{3}=0$. This is in contradiction with the assumption of $b_{2}$, $b_{3}\neq0$.

\textbf{Case 4.} $|\Psi_{0}\rangle$ consists of four terms, that is, $|\Psi_{0}\rangle=a_{0}|00\rangle+a_{1}|01\rangle+a_{2}|10\rangle+a_{3}|11\rangle$, where $a_{0}$, $a_{1}$, $a_{2}$, $a_{3}\neq0$ and $\sum_{i=0}^{3}|a_{i}|^{2}=1$. Then we acquire
\begin{align*}
{\rm Tr}_{A}(|\Psi_{0}\rangle\langle\Psi_{0}|)=(|&a_{0}|^{2}+|a_{2}|^{2})|0\rangle\langle0|+(|a_{1}|^{2}+|a_{3}|^{2})|1\rangle\langle1|\nonumber\\+(&a_{0}a_{1}^{*}+a_{2}a_{3}^{*})|0\rangle\langle1|+a_{0}^{*}a_{1}|1
\rangle\langle0|,\\\\
{\rm Tr}_{B}(|\Psi_{0}\rangle\langle\Psi_{0}|)=(|&a_{0}|^{2}+|a_{1}|^{2})|0\rangle\langle0|+(|a_{2}|^{2}+|a_{3}|^{2})|1\rangle\langle1|\nonumber\\+(&a_{0}a_{2}^{*}+a_{1}a_{3}^{*})|0\rangle\langle1|+(a_{0}^{*}a_{2}+a_{1}^{*}a_{3})|1
\rangle\langle0|.
\end{align*}

Suppose that $|\Psi_{1}\rangle=b_{0}|i_{1}i_{2}\rangle+b_{1}|j_{1}j_{2}\rangle+b_{2}|k_{1}k_{2}\rangle+b_{3}|l_{1}l_{2}\rangle$, where $\sum_{i=0}^{3}|b_{i}|^{2}=1$, $b_{0}$, $b_{1}$, $b_{2}$, $b_{3}\neq0$ and $i_{1}$, $i_{2}$, $j_{1}$, $j_{2}$, $k_{1}$, $k_{2}$, $l_{1}$, $l_{2}\in\{0, 1\}$. We make a similar argument as the proof of Case 3 and acquire the Table 4.
\begin{table}[htbp]
 \centering
 \caption{$|\Psi_{0}\rangle$ consists of four terms(The duplicate parts have been deleted)}
  \label{table:4}
  \begin{tabular}{|c|c|c|c|c|}
  \hline
  \multicolumn{4}{|c|}{$|\Psi_{1}\rangle=b_{0}|i_{1}i_{2}\rangle+b_{1}|j_{1}j_{2}\rangle+b_{2}|k_{1}k_{2}\rangle+b_{3}|l_{1}l_{2}\rangle$} & \multirow{2}{*}{Masking or not}\\
  \cline{1-4}
  $|i_{1}i_{2}\rangle$ & $|j_{1}j_{2}\rangle$ & $|k_{1}k_{2}\rangle$ & $|l_{1}l_{2}\rangle$ & \\
  \hline
  \multirow{8}{*}{$|00\rangle$} & $|00\rangle$ & $|00\rangle$ & $|00\rangle$ & $\times$\\
  \cline{2-5}
  & $|00\rangle$ & $|00\rangle$ & $|01\rangle$ & $\times$\\
  \cline{2-5}
  & $|00\rangle$ & $|00\rangle$ & $|10\rangle$ & $\times$\\
  \cline{2-5}
  & $|00\rangle$ & $|00\rangle$ & $|11\rangle$ & $\times$\\
  \cline{2-5}
  & $|00\rangle$ & $|01\rangle$ & $|10\rangle$ & $\times$\\
  \cline{2-5}
  & $|00\rangle$ & $|01\rangle$ & $|11\rangle$ & $\times$\\
  \cline{2-5}
  & $|00\rangle$ & $|10\rangle$ & $|11\rangle$ & $\times$\\
  \cline{2-5}
  & $|00\rangle$ & $|10\rangle$ & $|11\rangle$ & $\times$\\
  \cline{2-5}
  & $|01\rangle$ & $|10\rangle$ & $|11\rangle$ &$\surd$\\
  \hline
  \multirow{3}{*}{$|01\rangle$}& $|01\rangle$ & $|01\rangle$ & $|01\rangle$ & $\times$\\
  \cline{2-5}
  & $|01\rangle$ & $|01\rangle$ & $|11\rangle$ & $\times$\\
  \cline{2-5}
  & $|01\rangle$ & $|10\rangle$ & $|11\rangle$ & $\times$\\
  \hline
  \multirow{2}{*}{$|10\rangle$} & $|10\rangle$ & $|10\rangle$ & $|10\rangle$ & $\times$\\
  \cline{2-5}
  & $|10\rangle$ & $|10\rangle$ & $|11\rangle$ & $\times$\\
  \hline
  $|10\rangle$ & $|10\rangle$ & $|10\rangle$ & $|10\rangle$ & $\times$\\
  \hline
  $|11\rangle$ & $|11\rangle$ & $|11\rangle$ & $|11\rangle$ & $\times$\\
  \hline
  \end{tabular}
\end{table}

Below, considering the case of $|\Psi_{1}\rangle=b_{0}|00\rangle+b_{1}|00\rangle+b_{2}|10\rangle+b_{3}|11\rangle$ in Table \ref{table:4}, we obtain
\begin{align*}
{\rm Tr}_{A}(|\Psi_{1}\rangle\langle\Psi_{1}|)=(&|b_{0}|^{2}+b_{0}b_{1}^{*}+b_{0}^{*}b_{1}+|b_{1}|^{2}+|b_{2}|^{2})|0\rangle\langle0|{}\nonumber\\{}+&|b_{3}|^{2}|1\rangle\langle1|+b_{2}b_{3}^{*}|0\rangle\langle1|+b_{2}^{*}b_{3}|1
\rangle\langle0|.
\end{align*}

Due to ${\rm Tr}_{A}(|\Psi_{0}\rangle\langle\Psi_{0}|)={\rm Tr}_{A}(|\Psi_{1}\rangle\langle\Psi_{1}|)$, we know that
\begin{align}
\begin{split}
\left \{
\begin{array}{llll}
|a_{0}|^{2}+|a_{2}|^{2}-|b_{0}|^{2}-b_{0}b_{1}^{*}-b_{0}^{*}b_{1}-|b_{1}|^{2}|b_{2}|^{2}=0\nonumber\\\\
|a_{1}|^{2}+|a_{3}|^{2}-|b_{3}|^{2}=0\nonumber\\\\
a_{0}a_{1}^{*}+a_{2}a_{3}^{*}-b_{2}b_{3}^{*}=0\nonumber\\\\
a_{0}^{*}a_{1}-b_{2}^{*}b_{3}=0.\nonumber
\end{array}
\right.
\end{split}
\end{align}

As a result, we acquire that $a_{2}=0$ or $a_{3}=0$, which contradicts the condition $a_{2}$, $a_{3}\neq0$.

In summary, the conclusion is correct, i.e., a single qubit state can be masked by two non-orthogonal quantum states, which have the same number of terms and the same basis.

\subsection{Masking of orthogonal quantum states}

In this section, we discuss the further optimization of the masking conditions when $|\Psi_{0}\rangle$ and $|\Psi_{1}\rangle$ are orthogonal.

Let $|\Psi_{0}\rangle=a_{0}|00\rangle+a_{1}|11\rangle$, $|\Psi_{1}\rangle=b_{0}|01\rangle+b_{1}|10\rangle$ and $|\Psi\rangle=\alpha_{0}|\Psi_{0}\rangle+\alpha_{1}|\Psi_{1}\rangle$, where $\sum_{i=0}^{1}|a_{i}|^{2}=\sum_{i=0}^{1}|b_{i}|^{2}=\sum_{i=0}^{1}|\alpha_{i}|^{2}=1$. Under this assumption, the masking conditions Eq.(\ref{4}), Eq.(\ref{7}) and Eq.(\ref{8}) can be further simplified as
\begin{align}
\begin{split}
\left \{
\begin{array}{lll}
|a_{0}|^{2}=|a_{1}|^{2}=|b_{0}|^{2}=|b_{1}|^{2}=\frac{1}{2}\\\\
\alpha_{0}\alpha_{1}^{*}a_{0}b_{0}^{*}+\alpha_{0}^{*}\alpha_{1}a_{1}^{*}b_{1}=0\\\\
\alpha_{0}\alpha_{1}^{*}a_{0}b_{1}^{*}+\alpha_{0}^{*}\alpha_{1}a_{1}^{*}b_{0}=0,
\end{array}
\right.
\end{split}
\label{9}
\end{align}
and the quantum states $|\Psi_{0}\rangle$, $|\Psi_{1}\rangle$ and $|\Psi\rangle$ satisfy the masking definition
\begin{align*}
{\rm Tr}_{A}(|\Psi\rangle\langle\Psi|)={\rm Tr}_{A}(|\Psi_{0}\rangle\langle\Psi_{0}|)={\rm Tr}_{A}(|\Psi_{1}\rangle\langle\Psi_{1}|)=\frac{|0\rangle\langle0|+|1\rangle\langle1|}{2},\\\\
{\rm Tr}_{B}(|\Psi\rangle\langle\Psi|)={\rm Tr}_{B}(|\Psi_{0}\rangle\langle\Psi_{0}|)={\rm Tr}_{B}(|\Psi_{1}\rangle\langle\Psi_{1}|)=\frac{|0\rangle\langle0|+|1\rangle\langle1|}{2}.
\end{align*}

We show that when the coefficients of $|\Psi_{0}\rangle$, $|\Psi_{1}\rangle$ and $|\Psi\rangle$ satisfy Eq.(\ref{9}), quantum information contained in the quantum state $|b\rangle=\alpha_{0}|0\rangle+\alpha_{1}|1\rangle$ can be masked. To demonstrate this claim, we consider the following two examples.

\textbf{Example 1.} For Eq.(\ref{9}), we analyze the case that $a_{0}$, $a_{1}$, $b_{0}$ and $b_{1}$ are given. Then we discuss what kind of state $|b\rangle=\alpha_{0}|0\rangle+\alpha_{1}|1\rangle$ can be masked by $|\Psi_{0}\rangle$ and $|\Psi_{1}\rangle$.

Given
\begin{align*}
|\Psi_{0}\rangle=\frac{1}{\sqrt{2}}|00\rangle+\frac{i}{\sqrt{2}}|11\rangle,\\
\\
|\Psi_{1}\rangle=\frac{1}{\sqrt{2}}|01\rangle+\frac{1}{\sqrt{2}}|10\rangle,
\end{align*}
we obtain
\begin{align*}
|\Psi\rangle=\alpha_{0}|\Psi_{0}\rangle+\alpha_{1}|\Psi_{1}\rangle=\alpha_{0}(\frac{1}{\sqrt{2}}|00\rangle+\frac{i}{\sqrt{2}}|11\rangle)+\alpha_{1}(\frac{1}{\sqrt{2}}|01\rangle+\frac{1}{\sqrt{2}}|10\rangle).
\end{align*}

Therefore, Eq.(\ref{9}) can be simplified to
\begin{align}
\frac{1}{2}\alpha_{0}\alpha_{1}^{*}-\frac{i}{2}\alpha_{0}^{*}\alpha_{1}=0.
\label{10}
\end{align}

Let $\alpha_{0}=x_{0}+y_{0}i$ and $\alpha_{1}=x_{1}+y_{1}i$. Due to $\sum_{i=0}^{1}|\alpha_{i}|^{2}=1$, we acquire
\begin{align*}
y_{1}=\pm\sqrt{1-(x_{0}^{2}+y_{0}^{2}+x_{1}^{2})},
\end{align*}
where $x_{0}$, $x_{1}$, $y_{0}\in[-1,1]$ and $x_{0}^{2}+y_{0}^{2}+x_{1}^{2}\in[0,1]$.

Then Eq.(\ref{10}) is reformulated as
\begin{align}
&x_{0}x_{1}+y_{0}\sqrt{1-(x_{0}^{2}+y_{0}^{2}+x_{1}^{2})}+x_{0}\sqrt{1-(x_{0}^{2}+y_{0}^{2}+x_{1}^{2})}-x_{1}y_{0}=0,\label{11}
\end{align}
or
\begin{align}
&x_{0}x_{1}-y_{0}\sqrt{1-(x_{0}^{2}+y_{0}^{2}+x_{1}^{2})}-x_{0}\sqrt{1-(x_{0}^{2}+y_{0}^{2}+x_{1}^{2})}-x_{1}y_{0}=0.\label{12}
\end{align}

\begin{figure}[htb]
\centering
\subfigure[]{
\includegraphics[width=6cm]{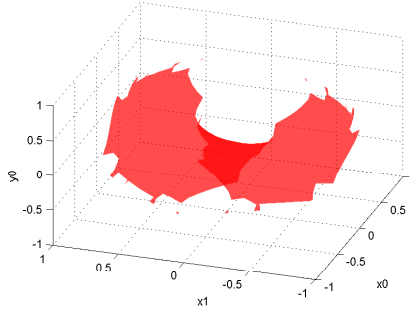}\label{fig1}
}

\subfigure[]{
\includegraphics[width=6cm]{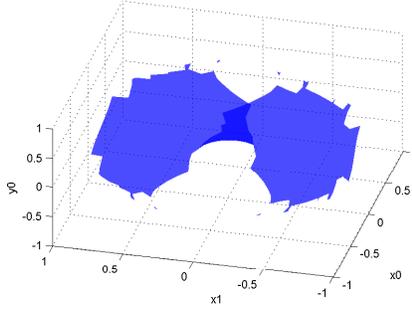}\label{fig2}
}
\caption{\subref{fig1} $|\Psi_{0}\rangle=\frac{1}{\sqrt{2}}|00\rangle+\frac{i}{\sqrt{2}}|11\rangle$ and $|\Psi_{1}\rangle=\frac{1}{\sqrt{2}}(|01\rangle+|10\rangle)$ can mask information in $|b\rangle=\alpha_{0}|0\rangle+\alpha_{1}|1\rangle$, if and only if $\alpha_{0}=x_{0}+y_{0}i$ and $\alpha_{1}=x_{1}+\sqrt{1-(x_{0}^{2}+y_{0}^{2}+x_{1}^{2})}i$ satisfy this figure. \subref{fig2} $|\Psi_{0}\rangle=\frac{1}{\sqrt{2}}|00\rangle+\frac{i}{\sqrt{2}}|11\rangle$ and $|\Psi_{1}\rangle=\frac{1}{\sqrt{2}}(|01\rangle+|10\rangle)$ can mask information in $|b\rangle=\alpha_{0}|0\rangle+\alpha_{1}|1\rangle$, if and only if $\alpha_{0}=x_{0}+y_{0}i$ and $\alpha_{1}=x_{1}-\sqrt{1-(x_{0}^{2}+y_{0}^{2}+x_{1}^{2})}i$ satisfy this figure.}
\end{figure}

As shown in Figure 1, we find a kind of states $|b\rangle=\alpha_{0}|0\rangle+\alpha_{1}|1\rangle$ which can be masked, i.e., when the coefficients of the state $|b\rangle$ satisfy the relationship in Figure 1, the state $|b\rangle=\alpha_{0}|0\rangle+\alpha_{1}|1\rangle$ can be masked by given quantum states $|\Psi_{0}\rangle=\frac{1}{\sqrt{2}}|00\rangle+\frac{i}{\sqrt{2}}|11\rangle$ and $|\Psi_{1}\rangle=\frac{1}{\sqrt{2}}|01\rangle+\frac{1}{\sqrt{2}}|10\rangle$.

In particular, according to Figure 1\subref{fig2}, we consider the case of $x_{0}=-\frac{1}{5}$, $x_{1}=0$ and $y_{0}=-\frac{1}{5}$, namely, $\alpha_{0}=\frac{1-i}{5}$ and $\alpha_{1}=\frac{\sqrt{23}}{5}$.

Through calculation, we obtain
\begin{align*}
{\rm Tr}_{A}(|\Psi_{0}\rangle\langle\Psi_{0}|)={\rm Tr}_{A}(|\Psi_{1}\rangle\langle\Psi_{1}|)={\rm Tr}_{A}(|\Psi\rangle\langle\Psi|)=\frac{|0\rangle\langle0|+|1\rangle\langle1|}{2},\\
{\rm Tr}_{B}(|\Psi_{0}\rangle\langle\Psi_{0}|)={\rm Tr}_{B}(|\Psi_{1}\rangle\langle\Psi_{1}|)={\rm Tr}_{B}(|\Psi\rangle\langle\Psi|)=\frac{|0\rangle\langle0|+|1\rangle\langle1|}{2}.
\end{align*}

In other words, we know that the state $|b\rangle=\frac{1-i}{5}|0\rangle+\frac{\sqrt{23}}{5}|1\rangle$ can be masked by the given quantum states $|\Psi_{0}\rangle=\frac{1}{\sqrt{2}}|00\rangle+\frac{i}{\sqrt{2}}|11\rangle$ and $|\Psi_{1}\rangle=\frac{1}{\sqrt{2}}|01\rangle+\frac{1}{\sqrt{2}}|10\rangle$.

\textbf{Example 2.} By Eq.(\ref{9}), we consider that if $\alpha_{0}$ and $\alpha_{1}$ are certain, what kind of quantum states $|\Psi_{0}\rangle$ and $|\Psi_{1}\rangle$ can mask the quantum information in $|b\rangle=\alpha_{0}|0\rangle+\alpha_{1}|1\rangle$.

Let $\alpha_{0}=\frac{1}{\sqrt{1+\lambda^{2}}}$ and $\alpha_{1}=(\frac{\lambda}{\sqrt{1+\lambda^{2}}})i$, where $\lambda\in\mathbb{R}$, and denote
\begin{align*}
a_{0}=x_{0}+y_{0}i,\qquad a_{1}=x_{1}+y_{1}i,\\
b_{0}=x_{2}+y_{2}i,\qquad b_{1}=x_{3}+y_{3}i.
\end{align*}
Then, Eq.(\ref{9}) can be reduced to
\begin{align}
\begin{split}
\left \{
\begin{array}{lllll}
x_{0}^{2}+y_{0}^{2}=x_{1}^{2}+y_{1}^{2}=x_{2}^{2}+y_{2}^{2}=x_{3}^{2}+y_{3}^{2}=\frac{1}{2}\\\\
-x_{0}x_{2}+x_{3}x_{1}-y_{0}y_{2}+y_{3}y_{1}=0\\\\
-x_{0}x_{3}+x_{2}x_{1}-y_{0}y_{3}+y_{2}y_{1}=0\\\\
-x_{0}y_{2}+x_{2}y_{0}+x_{3}y_{1}-x_{1}y_{3}=0\\\\
x_{3}y_{0}-x_{0}y_{3}+x_{2}y_{1}-x_{1}y_{2}=0.
\end{array}
\right.
\end{split}
\label{13}
\end{align}

Particularly, when $x_{0}=x_{2}$ and $y_{0}=y_{2}$, Eq.(\ref{13}) is further changed to
\begin{align}
\begin{split}
\left \{
\begin{array}{lll}
x_{3}x_{1}+y_{3}y_{1}-\frac{1}{2}=0\\\\
x_{3}y_{1}-x_{1}y_{3}=0\\\\
(x_{0}+y_{0})(x_{3}-x_{1})+(y_{0}-x_{0})(y_{3}-y_{1})=0.
\end{array}
\right.
\end{split}
\label{14}
\end{align}

Since $x_{i}^{2}+y_{i}^{2}=\frac{1}{2}$, where $i\in\{0,1,2,3\}$, denoting
\begin{align*}
\begin{split}
\left \{
\begin{array}{llll}
y_{0}=\pm\sqrt{\frac{1}{2}-x_{0}^{2}}\\\\
y_{1}=\pm\sqrt{\frac{1}{2}-x_{1}^{2}}\\\\
y_{2}=y_{0}=\pm\sqrt{\frac{1}{2}-x_{0}^{2}}\\\\
y_{3}=\pm\sqrt{\frac{1}{2}-x_{3}^{2}}.
\end{array}
\right.
\end{split}
\end{align*}

Through the above equations, we get the solutions to Eq.(\ref{14}), that is, $x_{1}=x_{3}=\frac{\sqrt{2}}{2}$, $x_{0}\in\mathbb{R}$ or $x_{1}=x_{3}=-\frac{\sqrt{2}}{2}$, $x_{0}\in\mathbb{R}$. Indeed, we obtain that
\begin{align}
\begin{split}
\left \{
\begin{array}{ll}
|\Psi_{0}\rangle=(x_{0}+y_{0}i)|00\rangle+\frac{\sqrt{2}}{2}|11\rangle\\\\
|\Psi_{1}\rangle=(x_{0}+y_{0}i)|01\rangle+\frac{\sqrt{2}}{2}|10\rangle,
\end{array}
\right.
\end{split}
\label{15}
\end{align}
or
\begin{align}
\begin{split}
\left \{
\begin{array}{ll}
|\Psi_{0}\rangle=(x_{0}+y_{0}i)|00\rangle-\frac{\sqrt{2}}{2}|11\rangle\\\\
|\Psi_{1}\rangle=(x_{0}+y_{0}i)|01\rangle-\frac{\sqrt{2}}{2}|10\rangle.
\end{array}
\right.
\end{split}
\label{16}
\end{align}

According to Eq.(\ref{15}) or Eq.(\ref{16}), Eq.(\ref{9}) is simplified as
\begin{align}
(-x_{0}^{2}-y_{0}^{2}+\frac{1}{2})\frac{\lambda}{1+\lambda^{2}}=0.
\label{17}
\end{align}
Moreover, let $\lambda\in[-1,1]$, then we draw the Figure 2 for an intuitive description.
\begin{figure}[htb]
\centering
\includegraphics[width=8cm]{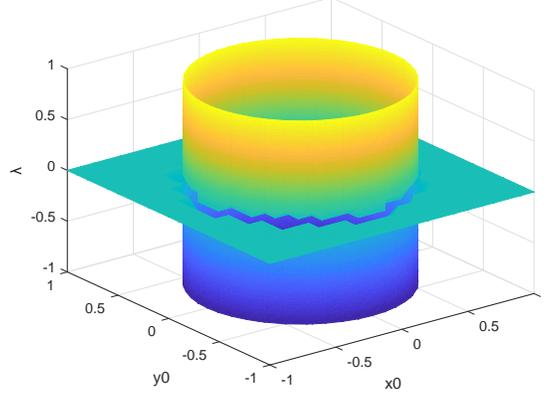}
\caption{$|b\rangle=(\frac{1}{\sqrt{1+\lambda^{2}}})|0\rangle+(\frac{\lambda i}{\sqrt{1+\lambda^{2}}})|1\rangle$ can be masked by $|\Psi_{0}\rangle=(x_{0}+y_{0}i)|00\rangle+\frac{\sqrt{2}}{2}|11\rangle$ and $|\Psi_{1}\rangle=(x_{0}+y_{0}i)|01\rangle+\frac{\sqrt{2}}{2}|10\rangle$ or $|\Psi_{0}\rangle=(x_{0}+y_{0}i)|00\rangle-\frac{\sqrt{2}}{2}|11\rangle$ and $|\Psi_{1}\rangle=(x_{0}+y_{0}i)|01\rangle-\frac{\sqrt{2}}{2}|10\rangle$, if and only if their coefficients $\lambda$, $x_{0}$, $y_{0}$ satisfy the relationship represented in this image.}
\end{figure}

As shown in Figure 2, for the state $|b\rangle=(\frac{1}{\sqrt{1+\lambda^{2}}})|0\rangle+(\frac{\lambda i}{\sqrt{1+\lambda^{2}}})|1\rangle$, there are quantum states $|\Psi_{0}\rangle$ and $|\Psi_{1}\rangle$ in Eq.(\ref{15}) or Eq.(\ref{16}), which can mask the state $|b\rangle$, if and only if their coefficients satisfy the relationship shown in Figure 2.

\section{Conclusion}\label{section3}
In two-dimensional Hilbert space, we expressed the quantum information masking conditions by a system of equations about the coefficients of quantum states. Moreover, by observing the characteristic of the maskable non-orthogonal two-qubit quantum states, we obtained the conclusion that if two non-orthogonal quantum states can mask a single qubit state, they have the same number of terms and the same basis. Furthermore, we considered two examples of orthogonal quantum states and calculated the masking conditions. Finally, we gave the corresponding images for an intuitive description. Our results would provide a foundation idea for further study on the masking of higher dimensional quantum states.

\section*{Acknowledgements}
    This work was supported by the Natural Science Foundation of Hebei Province(Grant No. A2019210057).

\end{document}